\documentclass{article}

\usepackage{amsthm,amsmath,amsfonts}
\usepackage{nameref}
\usepackage{hyperref}
\usepackage[utf8]{inputenc} %unicode support
\usepackage{graphicx}
\usepackage{authblk}

\newcommand{\bs}{\boldsymbol}
\newcommand{\bsxi}{{\bs \xi}}
\newcommand{\ph}{\hat{p}}
\newcommand{\Var}{\text{Var}}
\newcommand{\MC}[2]{\multicolumn{#1}{c}{#2}}

\begin{document}
\title{A global sensitivity analysis approach for morphogenesis models}
\author[1,2]{Sonja E.M. Boas\thanks{boas@cwi.nl}}
\author[3]{Maria I. Navarro Jimenez\thanks{mariaisabel.navarrojimenez@kaust.edu.sa}}
\author[1,2]{Roeland M.H. Merks\thanks{merks@cwi.nl}}
\author[1]{Joke G. Blom\thanks{joke.blom@cwi.nl}}
\affil[1]{Life Sciences, CWI, Science Park 123, 1098XG Amsterdam, The Netherlands}                                   
\affil[2]{Mathematical Institute, Leiden University, Niels Bohrweg 1, 2333CA Leiden, The Netherlands}
\affil[3]{CEMSE Division, King Abdullah University of Science and Technology (KAUST), Thuwal, Kingdom of Saudi Arabia}
\maketitle
\begin{abstract} % abstract
Morphogenesis is a developmental process in which cells organize into shapes and patterns. Complex, multi-factorial models are commonly used to study morphogenesis. It is difficult to understand the relation between the uncertainty in the input and the output of such `black-box' models, giving rise to the need for sensitivity analysis tools. In this paper, we introduce a workflow for a global sensitivity analysis approach to study the impact of single parameters and the interactions between them on the output of morphogenesis models. To demonstrate the workflow, we used a published, well-studied model of vascular morphogenesis. The parameters of the model represent cell properties and behaviors that drive the mechanisms of angiogenic sprouting. The global sensitivity analysis correctly identified the dominant parameters in the model, consistent with previous studies. Additionally, the analysis provides information on the relative impact of single parameters and of interactions between them. The uncertainty in the output of the model was largely caused by single parameters. The parameter interactions, although of low impact, provided new insights in the mechanisms of \emph{in silico} sprouting. Finally, the analysis indicated that the model could be reduced by one parameter.
We propose global sensitivity analysis as an alternative approach to study and validate the mechanisms of morphogenesis. Comparison of the ranking of the impact of the model parameters to knowledge derived from experimental data and validation of manipulation experiments can help to falsify models and to find the operand mechanisms in morphogenesis. The workflow is applicable to all `black-box' models, including high-throughput \emph{in vitro} models in which an output measure is affected by a set of experimental perturbations. 
\end{abstract}

\section*{Background}
Morphogenesis, the organization of multiple cells into shapes and patterns, is a key process in biological development. Computational modeling is commonly used to study mechanistic hypotheses on morphogenesis \cite{Tanaka2015,Merks2009,Iber2013,Iber2013_1,Boehm2010, Anderson2007,Herrero2009} as they allow for simplification and isolation of the process. These computational studies typically involve multi-scale, non-linear and multi-factorial models. So far, the behavior of these computational models is studied for one or occasionally two parameters at a time,
which can lead to a wrong interpretation for non-linear models. Studying all parameters collectively with global sensitivity analysis resolves this problem. 

In this paper, we introduce a workflow that uses global sensitivity analysis to find the relevant single parameters and parameter interactions in `black-box' models of morphogenesis, which are strongly non-linear and multifactorial. 
Sensitivity analyses of computational models enable us to identify the effects of uncertainties in parameter values on the model output. Local sensitivity analysis investigates the behavior of the model in a small region around the 
nominal parameter values and is most often used to study model robustness.
Global sensitivity analysis (GSA) covers the entire input parameter space, or a specifically selected region 
hereof. In its most powerful form, it gives information on the impact of individual parameters and  combinations thereof on a nonlinear model  for  arbitrary parameter distributions. This is what e.g. variance-based methods like FAST \cite{FAST99} and the Sobol' method \cite{Sobol90, Sobol01} do. This, of course, is computationally expensive, therefore many methods have been proposed with simplifying assumptions like linearity of the model (MLR \cite{Ostrom90}); methods that produce less sophisticated results, e.g. partial or no information on interactions (Morris method\cite{Morris91};\cite{Zheng06,Cho03}); are less robust like DGSM (\cite{DGSM09,SobolKucherenko09}); or that use prior knowledge of the model, like Bayesian DGSM \cite{GSABangaDoyle12}). In this paper we use the Sobol' method \cite{Sobol01}, where we have modified the original method for efficiency reasons (for more details see Section \textit{\nameref{method:GSA}}).
Moreover, we introduce an approach to determine the sampling size  \emph{a priori} with an \emph{a posteriori} error check. Thus, it is not likely that the proposed GSA will excel in computational efficiency, but it will excel in predictability of the costs and reliability of the results.
In the biological field GSA is mostly applied to models consisting of ordinary differential equations, e.g., in pharmacology \cite{GSApharma15,SobolPKPD15}, neurodynamics \cite{Aldemar15}, or gene expression \cite{GSAgenexoAy10} and biochemical pathways \cite{GSABangaDoyle12} in cells.

GSA can give interesting new insights into models of morphogenesis. Firstly, GSA predicts which parameters can best be tuned to affect the model output. When the model parameters can be associated with biological cell properties, extracellular matrix properties, or gene 
expression, knowledge of their influence on morphogenesis can give predictions for \emph{in vitro} perturbation experiments, e.g. genetic knock-outs. Secondly, apart from identifying the impact of single parameters, GSA notably identifies parameter interactions. These can give new mechanistic insights in the functioning of the model. Thirdly, GSA is a tool to reduce the number of parameters in the model. When the analysis indicates that parameter variation does not impact the model output, the parameter value can be fixed. Fourthly, GSA can be used to make a selection of models that support biologically plausible hypotheses in a set of contradicting mechanistic hypotheses.

As a case study, we performed GSA on a previously published \cite{Merks2008}, well-studied computational model of vascular morphogenesis. In the model, a spheroid of cells develops into a vascular network. Cells secrete a compound to which cells chemotact by migrating towards higher concentrations of the compound. Vascular networks form when chemotaxis is inhibited at cell-cell interfaces. Vascular endothelial growth factor (VEGF) is a candidate for the secreted compound and extensions of pseudopods in the direction of cell-cell contacts might be locally inhibited by interference of vascular endothelial cadherins with VEGF receptor 2 signaling. Because of the key role of such `contact-inhibited chemotaxis' in this model, we will henceforth refer to it as the `contact inhibition model'. Numerous alternative hypotheses for vascular morphogenesis have been proposed \cite{Oers2014,Szabo2010,Merks2006,KohnLuque2011,Merks2008,Bauer2007,Shirinifard2009,Merks2009,Czirok2015}, and it is unsure which of these - if any 
- is correct. Thus the contact inhibition model is here used as an example model for 
morphogenesis, while the proposed GSA approach can assist in falsifying mechanisms in the future.

  \begin{figure}
 \includegraphics[width=\textwidth]{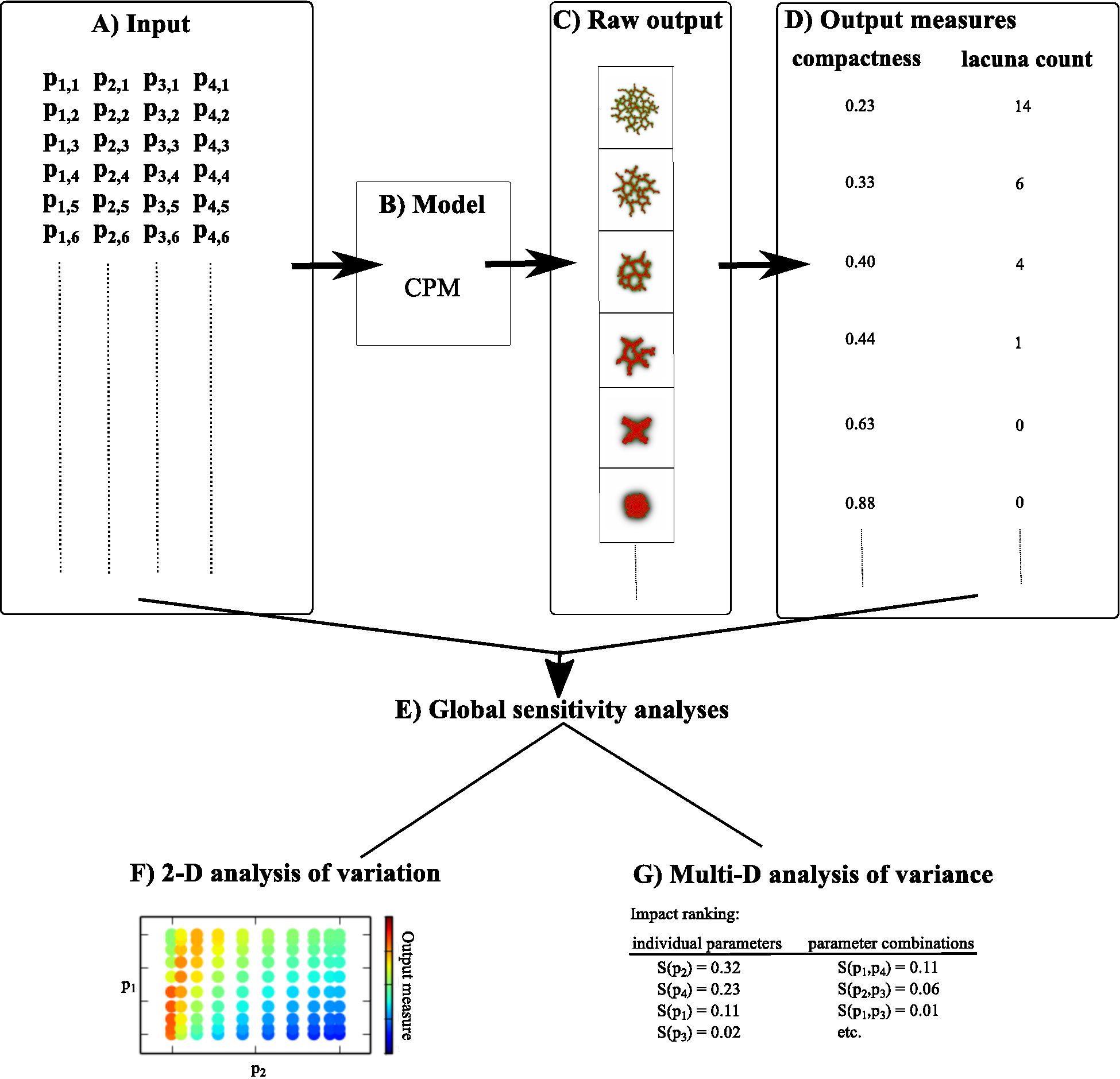}
  \caption{\textbf{Overview of the global sensitivity analysis.}
(A) The input of the model is a list of parameter sets. Each parameter set contains uniformly randomly selected values of parameters $p_{1}$ to $p_{4}$. This input is then fed into (B) the Cellular Potts Model (CPM)-based contact inhibition model. (C) The raw output of these models are images of the cell configuration at the end of the simulations. (D) Two output measures, compactness and lacuna count, are derived from these images. Two types of global sensitivity analysis are performed on these output measures (E). Firstly, intensity plots are used to study the impact of two-parameter combinations on the variation in the output measures (F). Secondly, Sobol' indices
are used to rank the impact of individual parameters and of parameter combinations on the variance of the output measures (G).}
\label{fig:workflow}
      \end{figure}
      
Figure \ref{fig:workflow} shows the workflow of the GSA analysis proposed in this paper. The input (Figure \ref{fig:workflow}A), a list of parameter sets, is fed into the cellular Potts-based contact inhibition model (Figure \ref{fig:workflow}B). This model generates images (Figure \ref{fig:workflow}C) of the resulting cell configuration as raw output, ranging from spheroids, to networks, to dispersed cells. Subsequently, two quantitative output measures (compactness and lacuna count) are derived from these images (Figure \ref{fig:workflow}D). Two types of GSA are performed on the output measures (Figure \ref{fig:workflow}E). Firstly, intensity plots show the impact of parameter combinations on the variation in the output measures (Figure \ref{fig:workflow}F). This analysis only allows for a two-dimensional GSA, in which the value of two parameters are varied simultaneously while keeping all other parameter values fixed. 
Secondly, a truly multivariate GSA ranks the impact of individual parameters and of parameter combinations on the variance of the output measures (Figure \ref{fig:workflow}G). Important aspects we address in this paper
are the reliability and the pitfalls of GSA.

\section*{Methods}\label{methods}
\subsection*{Vascular morphogenesis model}
The contact inhibition model \cite{Merks2008} is based on the Cellular Potts Model (CPM) \cite{Glazier1993,Graner1992}. Cells are projected on a regular square lattice ($\Lambda \subset \mathbb{Z}^2$) as patches of connected lattice sites, $\vec{x}$. Each lattice site, $\vec{x}\in\Lambda$, that is part of a cell is marked with that cell's identifier ($\sigma(\vec{x})$) and cell-free lattice sites represent extracellular matrix (ECM) with $\sigma=0$. Each cell identifier is associated with a type: $\tau(\sigma)\in\{\mathrm{ECM}, \mathrm{cell}, \mathrm{border}\}$. Cells have cell properties and behaviors, such as adhesion, cell size, or chemotaxis. The forces resulting from the biophysical properties of the cells and their active behavior are represented in the Hamiltonian ($H$) of the system 
\begin{equation}
H=\sum_{(\vec{x},\vec{x}\prime)}J(\tau(\sigma(\vec{x})),\tau(\sigma(\vec{x}\prime)))\left(1- \delta(\sigma(\vec{x}),\sigma(\vec{x}\prime))\right)+\lambda_{A}(\sigma)\sum_\sigma \left(A(\sigma)- a(\sigma)\right)^2,
\end{equation}
in which adhesion ($J$) is restricted to the cell membrane by the Kronecker delta ($\delta(x,y)=\{1,x=y;0,x\neq y\}$) and $(\vec{x},\vec{x}\prime)$ represents the set of adjacent lattice site pairs. There are three non-zero types of adhesion: $J_{\mathrm{cell},\mathrm{cell}}$, $J_{\mathrm{cell},\mathrm{border}}$ and $J_{\mathrm{cell},\mathrm{ECM}}$. $J_{\mathrm{cell},\mathrm{cell}}$ represents the adhesion strength between cells, and $J_{\mathrm{cell},\mathrm{ECM}}$ the adhesion strength of cells to the ECM. The lattice is surrounded with a border by 
which cells are repulsed, by setting $J_{\mathrm{cell},\mathrm{border}}=100$. The second term constrains the volume of cells, with $A$ representing the preferred size of a cell and $\lambda_{A}$ the rigidity of the cell. Deviation of the actual size ($a$) of cells from their preferred value increases the Hamiltonian. 

A cell moves by copying the state ($\sigma$) of a randomly selected lattice site $\vec{x}$ into a randomly selected adjacent lattice site $\vec{x}\prime$. In this manuscript, we use the eight, second order neighbors. These copies represent extensions and retractions of pseudopods at the cell membrane. A copy attempt that diminishes the Hamiltonian represents a move along a force and is always accepted. If a copy increases the Hamiltonian, it will only occur due to active movements of the cell membrane. We assume that such active motions are distributed according to a Boltzmann probability function: $P_{\mathrm{Boltzmann}}(H)=e^{\frac{-\Delta H}{\mu}}$, with $\mu$ a measure of the amplitude of random membrane fluctuations; $\mu$ is a scaling factor of the weights ($\lambda$) of the constraints and we fixed it at $\mu=50$ conform the settings in our previous work \cite{Merks2008}. The parameter values can only be qualitatively coupled to biological data.

We assume that cells secrete a chemoattractant at rate $\alpha$ ($s^{-1}$), producing a concentration field $c(\vec{x})$. The chemoattractant diffuses with a diffusion coefficient $D$ ($m^2/s$) and decays with rate $\epsilon$ ($s^{-1}$) in the ECM, resulting in the following partial differential equation (PDE): 
\begin{equation}\label{eq:PDE}
\frac{\partial c}{\partial t}=\alpha(1-\delta(\sigma(\vec{x}),0))-\epsilon \delta(\sigma(\vec{x}),0)c+D\nabla^{2}c, 
\end{equation}
such that secretion is located at the cells, where $\delta(\sigma(\vec{x}),0)=0$, and decay in the ECM. The field of the chemoattractant is initialized as $c(\vec{x})=0$ and fixed boundary conditions are imposed.
Cells can respond to this chemoattractant by migrating towards higher concentrations (chemotaxis). To this end, the change in the Hamiltonian by that copy, $\Delta H$, is augmented with $\Delta H_{\mathrm{chemotaxis}}=\lambda_{c}(c(\vec{x})-c(\vec{x}\prime))$  \cite{Savill1997}, and contact-inhibition is implemented by setting $\lambda_{c}=0$ at cell-cell interfaces such that chemotaxis only occurs at the cell-ECM interface. 

During one time step, referred to as a Monte Carlo Step (MCS), as many copy attempts are performed as there are sites in the lattice. The PDE for chemoattractant diffusion and degradation (Equation \ref{eq:PDE}) is discretized on the CPM lattice and we solve it numerically using a finite-difference scheme. We use 15 diffusion steps per MCS. The model is initialized with a centralized spheroid of 256 cells within a lattice of 400*400 sites (lattice spacing $2 \mu m$). We run the model for 5000 MCS, each representing 30 seconds, as networks are well formed in this time in the model as well as \emph{in vitro} \cite{Merks2006}.  

 \begin{figure}
  \includegraphics[width=\textwidth]{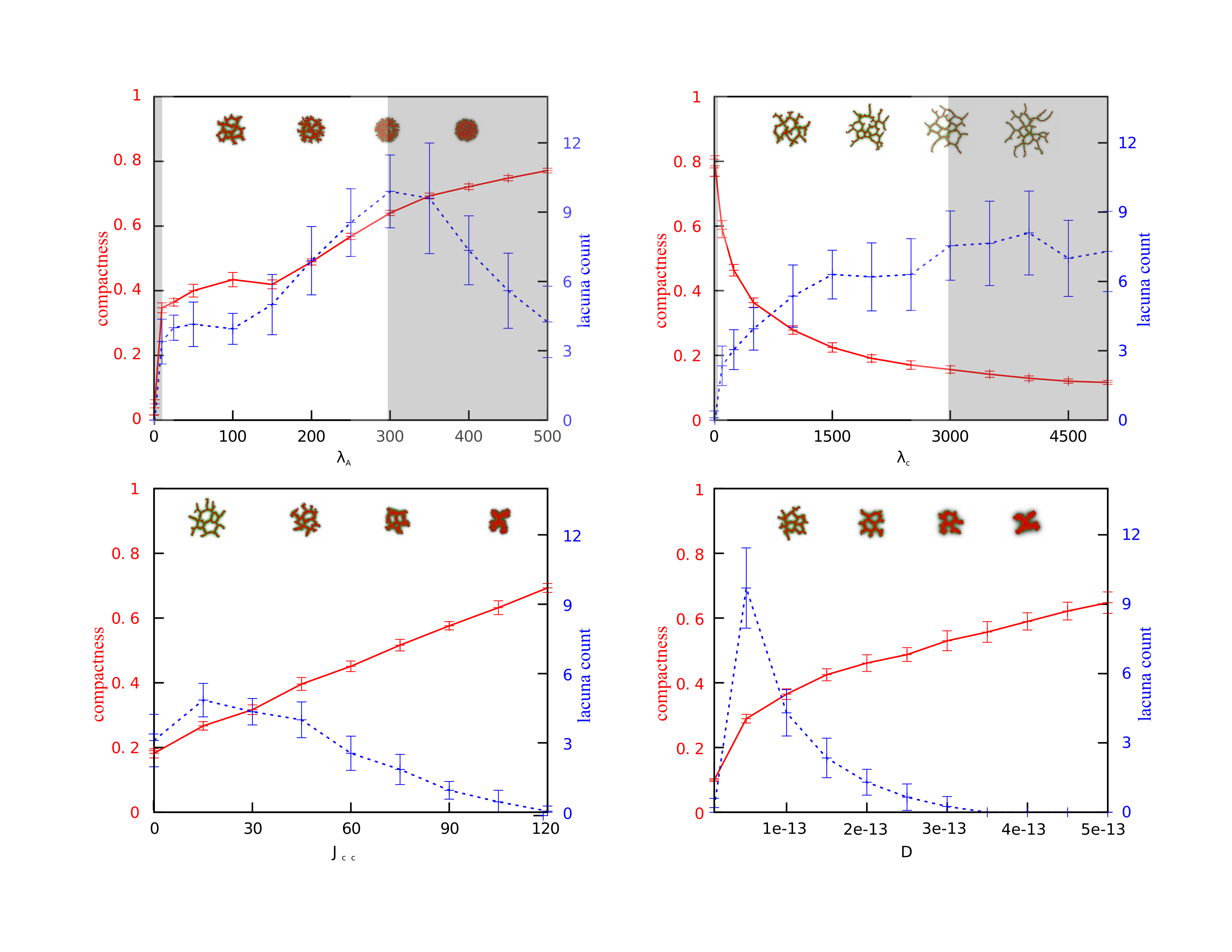}
  \caption{\textbf{One-dimensional parameter sweeps for compactness and lacuna count.} Plots of
  one-dimensional parameter sweeps for each of the four selected parameters: cell rigidity ($\lambda_A$), cell-cell adhesion ($J_{\mathrm{cell},\mathrm{cell}}$), diffusion coefficient of the chemoattractant ($D$), and sensitivity of cells to the chemoattractant at cell-matrix interfaces ($\lambda_c$). The red lines indicate compactness and the blue lines lacuna count (mean and standard deviation of 20 simulations).   }\label{fig:1DSA}
	\end{figure}

 \begin{figure}
   \includegraphics[width=\textwidth]{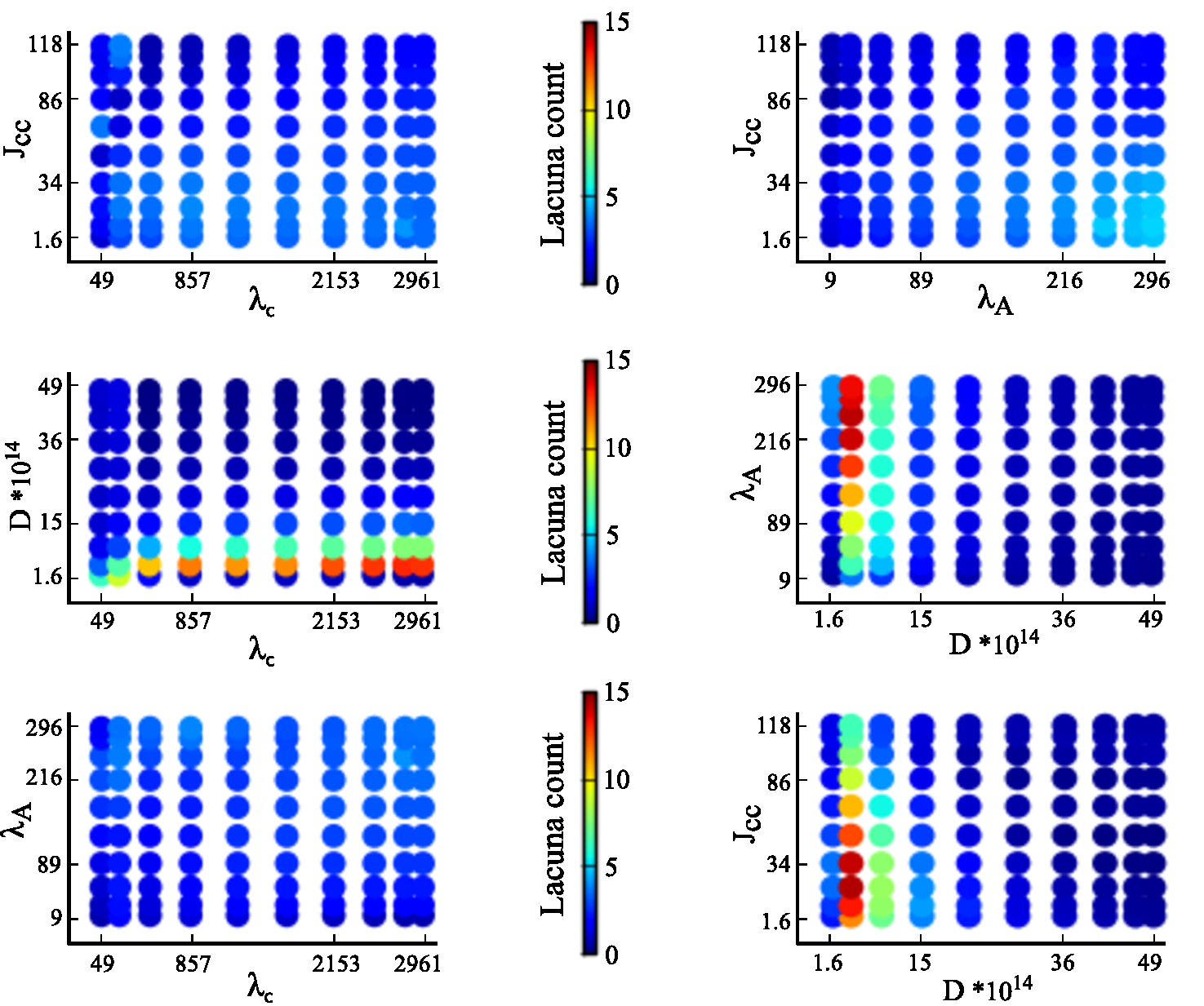}
  \caption{\textbf{Two-dimensional intensity plots of lacuna count.}
  The intensity of the output measure lacuna count, mapped to an interval of 0 to 15 as indicated by the color bars, is plotted for each two-parameter combination of the parameters cell rigidity ($\lambda_A$), cell-cell adhesion ($J_{\mathrm{cell},\mathrm{cell}}$), diffusion coefficient of the chemoattractant ($D$), and sensitivity of cells to the chemoattractant at cell-matrix interfaces ($\lambda_c$).    }\label{fig:2DSALC}
	\end{figure}

\begin{figure}
  \includegraphics[width=\textwidth]{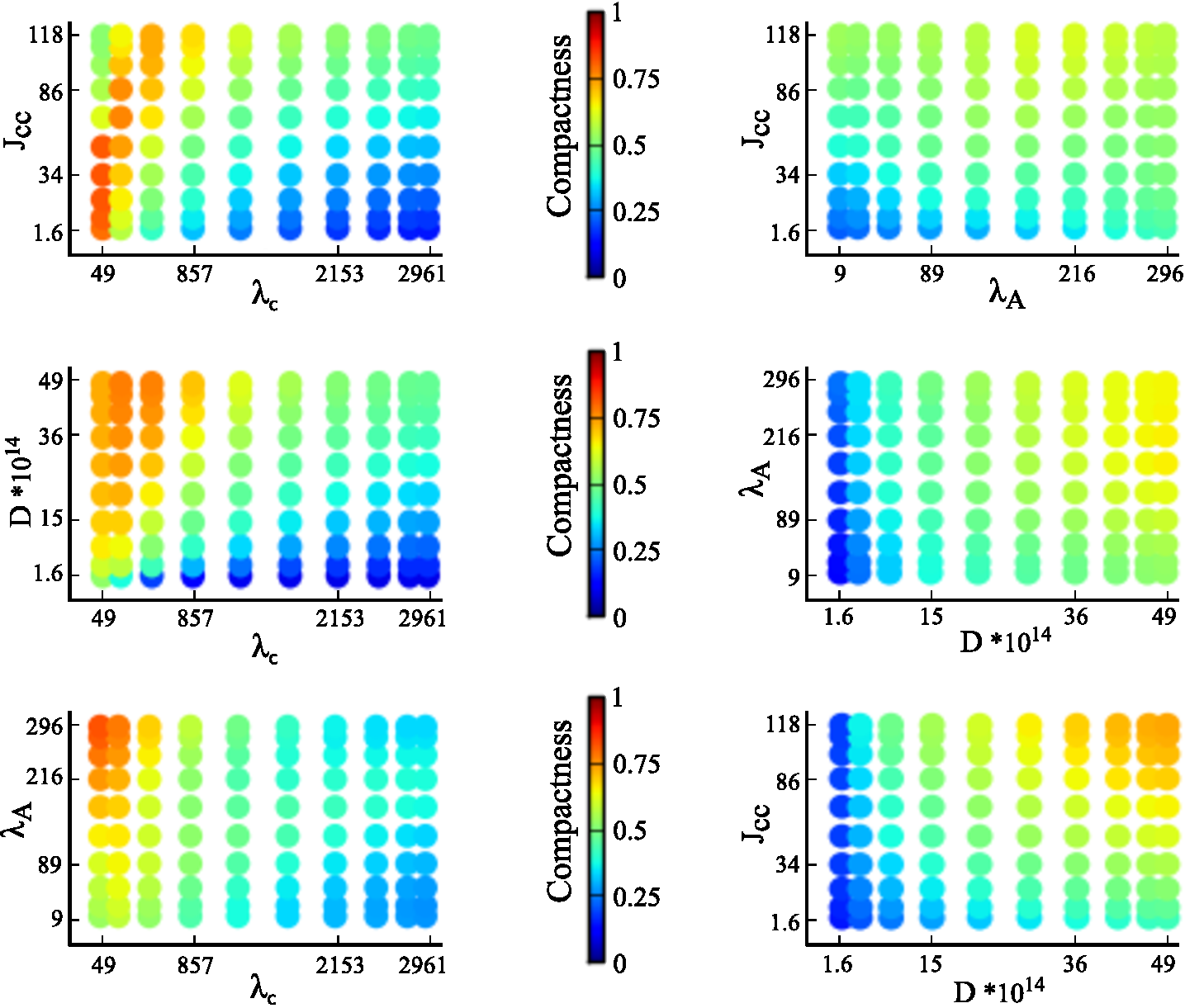}
  \caption{\textbf{Two-dimensional intensity plots of compactness.}
 The intensity of the output measure compactness, mapped to an interval of 0 to 1 as indicated by the color bars, is plotted for each two-parameter combination of the parameters cell rigidity ($\lambda_A$), cell-cell adhesion ($J_{\mathrm{cell},\mathrm{cell}}$), diffusion coefficient of the chemoattractant ($D$), and sensitivity of cells to the chemoattractant at cell-matrix interfaces ($\lambda_c$).    }\label{fig:2DSACmp}
	\end{figure}
	
\subsection*{Global sensitivity analysis}\label{method:GSA}
The variation in a solution or a measure thereof, like compactness, over the complete parameter space can only be visually inspected by looking at one or at most two parameters at a time while keeping the others fixed (cf. Figure~\ref{fig:1DSA},
\ref{fig:2DSALC}, and \ref{fig:2DSACmp}).  Since different measures will produce different multivariate output distributions and therefore also might result in a different outcome of the GSA, it is important to choose a measure or, more likely a number of measures that are significant for the study at hand.
If one wants to take the influence of all parameters simultaneously into account some form of a global measure of the multivariate output distribution is required. One such a measure is the variance of the distribution, which will be used in this paper.

We are specifically interested whether parameter {\em interactions} have a large impact on the output of this specific CPM-based model. Interactions of the parameters are unpredictable in non-linear models such as the CPM, but their impact is significant, since a large combined effect of parameters on the output impedes the experimental testing of a predicted effect of a single parameter.

 Sobol' \cite{Sobol90,Sobol01} introduced so-called
global sensitivity indices that describe the impact of specific parameters or combinations thereof on the uncertainty
in the model output and more in particular on the variance of the output distribution, hence the term `variance-based'
GSA. In the original method the necessary integrals are computed with Monte Carlo. However, the Sobol' indices 
can be computed very efficiently when the distribution of the output measure or response surface is expanded into a 
series of orthogonal polynomials, the Polynomial Chaos Expansion (PCE) \cite{Wiener38,XiuRevgPCE}. 
An overview of this method, using the same notation as in the following, can be found in \cite{gPCcor14};
here we summarize only the most important definitions for the case that the stochastic
input consists of independently uniformly distributed random variables. 
Note, however, that the method can also be applied for arbitrary, even non-parametric, distributions, allowing for data-driven GSA (see also \cite{aPC}).
The strength of the  method described below is that it
(i) can efficiently study multiple output measures derived from the output images, 
(ii) can robustly identify parameter interactions, and
(iii) checks the reliability of the result.  
 
Let $\bsxi$ be the $n$-dimensional vector of the independently uniformly distributed input parameters and $\varrho(\bsxi)$ its joint probability density function (pdf).
The output measure $u(\bsxi)$, e.g., of the (black-box) Cellular Potts Model, is expanded into a 
truncated series of polynomials that are orthogonal with respect to the pdf $\varrho$, separating the output into a deterministic and a stochastic part
\begin{equation}\label{eq:uPCEapprox}
 u(\bsxi) \approx \sum_{i=0}^N u_i \Phi_i(\bsxi),
\end{equation}
where the $n$-variate polynomials $\Phi_i(\bsxi)$ are products of $n$ univariate Legendre polynomials. 
The number of expansion terms $N$ is given by  $N+1 = \frac{(n+\ph)!}{n!\ph!}$, with $n$ the number of parameters and 
the {\em approximation order} $\ph$ the highest order of $\Phi_i$. 

To compute the expansion coefficients $u_i$ of Equation (\ref{eq:uPCEapprox}) we apply {\em Spectral Projection} which has the advantage that it can be used for 
black-box models since it projects the solution - and not the model - %(\ref{eq:uPCE})  
onto the polynomial space
\begin{equation}
 u_i = \frac{\left<u(\bsxi),\Phi_i(\bsxi)\right>}{\left<\Phi_i(\bsxi),\Phi_i(\bsxi)\right>}=
 \frac{1}{||\Phi_i||^2}
 \int_\Xi u(\bsxi) \, \Phi_i(\bsxi) \, \varrho(\bsxi)\, \text{d}\bsxi,\qquad i=0,1, ..., N,
\label{eq:uiPCE}
\end{equation}
where $\Xi$ is the support of the joint pdf $\varrho(\bsxi)$. As the parameter inputs are independent,
both $\Phi_i$ and $\varrho$ can be written in product form; for the multivariate polynomial $\Phi_i$ this results in
a product of univariate polynomials
\begin{equation}
 \Phi_i(\bsxi) = \prod_{k=1}^n \Phi_{\textrm{index}(i,k)}(\xi_k), 
 \text{ with index}(i,k)=\{0,...,\ph\} \text{ and } \Phi_0(\xi_k)=1.
\end{equation}
The integrals in Equation (\ref{eq:uiPCE}) can then be computed by a repeated 
one-dimensional Gauss-Legendre quadrature rule
\begin{equation}\label{eq:uiPCEquad}
 u_i \approx \frac{1}{||\Phi_i||^2}\sum_{l_1=1}^{N_q} \cdots \sum_{l_n=1}^{N_q} 
 u(\xi_{l_1},\cdots,\xi_{l_n})\;\prod_{k=1}^n w_{l_k} \Phi_{\textrm{index}(i,k)}(\xi_{l_k}),
\end{equation}
with $N_q$ the number of quadrature points and $w$ the associated weights.
Note, that for integrals with a known weight function, like e.g. a pdf, 
Gauss quadrature has the optimal convergence order of $2N_q-1$ for $N_q$ quadrature points, where the points and the weights of the quadrature rule are dependent on
the weight function.

How to choose $N_q$ and $\ph$ to obtain reliable Sobol' indices will be the subject of Section  \textit{\nameref{method:reliable}}.

\subsubsection*{Statistics and Polynomial Chaos Expansion}
Using a PC expansion, the
only input needed to compute the moments and the Sobol' indices of the output distribution
are the expansion coefficients. E.g., the mean $\mu=u_0$ and the variance is given by
\begin{equation}\label{eq:PCEvar}
 \int_\Xi \left(u(\bsxi)-\mu\right)^2 \varrho(\bsxi) d\bsxi \approx
 \sum_{i=1}^N u^2_i||\Phi_i||^2 =: \Var_{\mathrm{PCE}}.
\end{equation}
Note, that the approximation, $\Var_{\mathrm{PCE}}$, is a monotonously increasing function of $N$ and thus of $\ph$.
The sum in the variance formula can be directly split into contributions from the various parameters or combinations thereof, 
the Sobol' indices (cf. \cite{Sudret08}). E.g., for the first-order Sobol' index for parameter $j$ only terms contribute if 
$\Phi_i(\bsxi)$ equals a univariate polynomial in $\xi_j$
\begin{equation}
 S_j \approx \frac{\sum_{i=1}^N  \textrm{bool}(i,j)\;u_i^2 ||\Phi_i||^2}{\Var_{\mathrm{PCE}}},
\end{equation}
where $\textrm{bool}(i,j)=\left( \textrm{index}(i,j) > 0 \wedge \textrm{index}(i,k) = 0, \forall k\neq j \right)$. 
For a combined influence of more than one parameter like $S_{13}$ the Sobol' index
can be computed analogously. The sum of all Sobol' indices equals one.

\subsection*{Reliable GSA in practice}\label{method:reliable}
At first sight the accuracy of the PCE approximation of the response surface - and thus of the 
statistics - seems to be determined by the {\em number} of expansion terms, $N$, in Equation (\ref{eq:uPCEapprox}). But the 
{\em accuracy} of the expansion coefficients $u_i$ also plays an important role. 
This accuracy is determined by the approximation (Equation \ref{eq:uiPCEquad}) of the integral in Equation (\ref{eq:uiPCE}), 
which is determined by the number of quadrature points, $N_q$. 
Moreover, the higher PCE order $\ph$ needed to obtain sufficient accuracy, the higher
the polynomial order of $\Phi_i(\bsxi)$ becomes, which increases the complexity of the integrand. 
If one computes the integral of a high 
order polynomial with a small amount of points, the resulting expansion coefficients are merely noise instead of being 
informative.

The question we want to answer in this section is how to determine the number of quadrature points, $N_q$, 
and the expansion order, $\ph$, to obtain a sufficiently high accuracy for
the coefficients to allow us to trust the Sobol' indices and more specifically the ranking of the parameters that
follows from it. 
Here, we sketch a method to determine the number of quadrature points. It relies on the fact that the Sobol'
indices are variance-based, i.e., one can not expect to compute Sobol' indices accurately from a PCE expansion for which the variance (Equation \ref{eq:PCEvar}) is not a sufficiently accurate approximation of the true value or at least 
comparable to the Gauss-Legendre quadrature approximation of the integral. So, let us define
\begin{equation}\label{eq:Errvar}
 err_{\Var} := \Var_{\mathrm{data}} - \Var_{\mathrm{PCE}},
\end{equation}
with 
\begin{eqnarray*}
 \Var_{\mathrm{data}} &=& 
 \sum_{l_1=1}^{N_q} \cdots \sum_{l_n=1}^{N_q} w_{l_1} \cdots w_{l_n} u(\xi_{l_1},\cdots \xi_{l_n})^2-\mu_{\mathrm{data}}^2,\\
 \mu_{\mathrm{data}}&=&\sum_{l_1=1}^{N_q} \cdots\sum_{l_n=1}^{N_q} w_{l_1} \cdots w_{l_n} u(\xi_{l_1},\cdots,\xi_{l_n}).
\end{eqnarray*}
For a given choice of $N_q$ one can easily compute PC expansions for various orders $\ph$. If $err_\Var$
is small and the required Sobol' indices have converged, the result can be trusted.

\begin{table}
\caption{ Statistics computed with insufficient  quadrature points. The resulting PCE 
approximation and thus the statistics can not be trusted.}
\begin{tabular}{r r r r r r r r r r}\hline
\MC{1}{$N_{q}$}&\MC{1}{$\hat{p}$}&\MC{1}{$\Var_{\mathrm{data}}$}&\MC{1}{$\Var_{\mathrm{PCE}}$}&\MC{1}{$S_1$}&\MC{1}{$S_2$}&\MC{1}{$S_{13}$}&\MC{1}{$S_3$}&\MC{1}{$S_{12}$}&\MC{1}{$S_{23}$}\\\hline
  \MC{1}{2}
&  1 &   4.09 &   4.09 &  1.00 &  0.00 &  0.00 &  0.00 &  0.00 &  0.00 \\ 
&  2 &   4.09 &   4.09 &  1.00 &  0.00 &  0.00 &  0.00 &  0.00 &  0.00 \\ 
&  3 &   4.09 &   8.32 &  1.00 &  0.00 &  0.00 &  0.00 &  0.00 &  0.00 \\ 
&  4 &   4.09 & 185.91 &  0.36 &  0.32 &  0.00 &  0.32 &  0.00 &  0.00 \\ 
&  5 &   4.09 & 197.44 &  0.34 &  0.30 &  0.03 &  0.30 &  0.03 &  0.00 \\\hline
  \MC{1}{5}
&  1 &  18.60 &   2.64 &  1.00 &  0.00 &  0.00 &  0.00 &  0.00 &  0.00 \\ 
&  2 &  18.60 &   2.70 &  0.98 &  0.02 &  0.00 &  0.00 &  0.00 &  0.00 \\ 
&  3 &  18.60 &   6.22 &  0.69 &  0.01 &  0.30 &  0.00 &  0.00 &  0.00 \\ 
&  4 &  18.60 &  17.17 &  0.25 &  0.64 &  0.11 &  0.00 &  0.00 &  0.00 \\ 
&  5 &  18.60 &  18.50 &  0.23 &  0.60 &  0.17 &  0.00 &  0.00 &  0.00 \\ 
&  6 &  18.60 &  29.49 &  0.14 &  0.75 &  0.11 &  0.00 &  0.00 &  0.00 \\ 
&  7 &  18.60 &  31.41 &  0.19 &  0.70 &  0.11 &  0.00 &  0.00 &  0.00 \\ 
&  8 &  18.60 &  31.50 &  0.19 &  0.70 &  0.11 &  0.00 &  0.00 &  0.00 \\ 
&  9 &  18.60 &  37.51 &  0.23 &  0.59 &  0.18 &  0.00 &  0.00 &  0.00 \\ 
& 10 &  18.60 &  88.82 &  0.29 &  0.43 &  0.08 &  0.20 &  0.00 &  0.00 \\\hline
\MC{2}{\ \  \ Exact}&\MC{2}{13.84}&0.31&0.44&0.24&\MC{1}{0}&\MC{1}{0}&\MC{1}{0}\\\hline
\end{tabular}

\label{tab:Ishigami1}
\end{table}

\begin{table}
\caption{Statistics computed with sufficient  quadrature points. The PCE approximation and the statistics show convergence (bold lines).}
\begin{tabular}{r r r r r r r r r r}\hline
\MC{1}{$N_{q}$}&\MC{1}{$\hat{p}$}&\MC{1}{$\Var_{\mathrm{data}}$}&\MC{1}{$\Var_{\mathrm{PCE}}$}&\MC{1}{$S_1$}&\MC{1}{$S_2$}&\MC{1}{$S_{13}$}&\MC{1}{$S_3$}&\MC{1}{$S_{12}$}&\MC{1}{$S_{23}$}\\\hline
  \MC{1}{8}
&  1 &  13.59 &   2.64 &  1.00 &  0.00 &  0.00 &  0.00 &  0.00 &  0.00 \\ 
&  2 &  13.59 &   3.00 &  0.88 &  0.12 &  0.00 &  0.00 &  0.00 &  0.00 \\ 
&  3 &  13.59 &   6.55 &  0.66 &  0.05 &  0.29 &  0.00 &  0.00 &  0.00 \\ 
&  4 &  13.59 &  10.34 &  0.42 &  0.40 &  0.18 &  0.00 &  0.00 &  0.00 \\ 
&  5 &  13.59 &  11.73 &  0.37 &  0.35 &  0.28 &  0.00 &  0.00 &  0.00 \\ 
&  6 &  13.59 &  13.45 &  0.32 &  0.44 &  0.24 &  0.00 &  0.00 &  0.00 \\ 
&\bf  7 &\bf  13.59 &\bf  13.59 &\bf  0.32 &\bf  0.43 &\bf  0.25 &\bf  0.00 &\bf  0.00 &\bf  0.00 \\ 
&\bf  8 &\bf  13.59 &\bf  13.59 &\bf  0.32 &\bf  0.43 &\bf  0.25 &\bf  0.00 &\bf  0.00 &\bf  0.00 \\ 
&\bf  9 &\bf  13.59 &\bf  13.59 &\bf  0.32 &\bf  0.43 &\bf  0.25 &\bf  0.00 &\bf  0.00 &\bf  0.00 \\ 
& 10 &  13.59 &  15.32 &  0.28 &  0.50 &  0.22 &  0.00 &  0.00 &  0.00 \\ 
& 11 &  13.59 &  15.35 &  0.29 &  0.49 &  0.22 &  0.00 &  0.00 &  0.00 \\ 
& 12 &  13.59 &  19.17 &  0.23 &  0.60 &  0.18 &  0.00 &  0.00 &  0.00 \\ 
& 13 &  13.59 &  21.08 &  0.29 &  0.54 &  0.17 &  0.00 &  0.00 &  0.00 \\ 
& 14 &  13.59 &  21.51 &  0.28 &  0.55 &  0.17 &  0.00 &  0.00 &  0.00 \\ 
& 15 &  13.59 &  27.62 &  0.32 &  0.43 &  0.25 &  0.00 &  0.00 &  0.00 \\\hline
  \MC{1}{10}
&  1 &  13.84 &   2.64 &  1.00 &  0.00 &  0.00 &  0.00 &  0.00 &  0.00 \\ 
&  2 &  13.84 &   3.00 &  0.88 &  0.12 &  0.00 &  0.00 &  0.00 &  0.00 \\ 
&  3 &  13.84 &   6.54 &  0.66 &  0.05 &  0.29 &  0.00 &  0.00 &  0.00 \\ 
&  4 &  13.84 &  10.36 &  0.42 &  0.40 &  0.18 &  0.00 &  0.00 &  0.00 \\ 
&  5 &  13.84 &  11.75 &  0.37 &  0.35 &  0.28 &  0.00 &  0.00 &  0.00 \\ 
&  6 &  13.84 &  13.59 &  0.32 &  0.44 &  0.24 &  0.00 &  0.00 &  0.00 \\ 
&  7 &  13.84 &  13.72 &  0.32 &  0.44 &  0.25 &  0.00 &  0.00 &  0.00 \\ 
&\bf  8 &\bf  13.84 &\bf  13.84 &\bf  0.31 &\bf  0.44 &\bf  0.24 &\bf  0.00 &\bf  0.00 &\bf  0.00 \\ 
&\bf  9 &\bf  13.84 &\bf  13.84 &\bf  0.31 &\bf  0.44 &\bf  0.24 &\bf  0.00 &\bf  0.00 &\bf  0.00 \\ 
&\bf 10 &\bf  13.84 &\bf  13.84 &\bf  0.31 &\bf  0.44 &\bf  0.24 &\bf  0.00 &\bf  0.00 &\bf  0.00 \\ 
&\bf 11 &\bf  13.84 &\bf  13.84 &\bf  0.31 &\bf  0.44 &\bf  0.24 &\bf  0.00 &\bf  0.00 &\bf  0.00 \\ 
& 12 &  13.84 &  13.95 &  0.31 &  0.45 &  0.24 &  0.00 &  0.00 &  0.00 \\ 
& 13 &  13.84 &  13.95 &  0.31 &  0.45 &  0.24 &  0.00 &  0.00 &  0.00 \\ 
& 14 &  13.84 &  15.81 &  0.27 &  0.51 &  0.21 &  0.00 &  0.00 &  0.00 \\ 
& 15 &  13.84 &  15.84 &  0.28 &  0.51 &  0.21 &  0.00 &  0.00 &  0.00 \\\hline
\MC{2}{\ \  \ Exact}&\MC{2}{13.84}&0.31&0.44&0.24&\MC{1}{0}&\MC{1}{0}&\MC{1}{0}\\\hline 
\end{tabular}

\label{tab:Ishigami2}
\end{table}

We illustrate this approach with a function for which the values of the statistics are analytically known, viz., the Ishigami 
function \cite{IshigamiHomma90, SobolLevitan99}
\begin{equation}\label{eq:Ishigami}
f(\bsxi) = \sin(\xi_1) + a \sin^2(\xi_2) + b\, \xi_3^4 \sin(\xi_1),
\end{equation}
with $\xi_i\sim\mathcal{U}[-\pi,\pi]$, $i=\{1,2,3\}$, and $a=7$ and $b=0.1$.
We compute the PCE approximation of Equation (\ref{eq:Ishigami}) for an increasing number of PCE terms and an increasing number 
of quadrature points. 
Table~\ref{tab:Ishigami1} illustrates the result of using not enough quadrature points ($N_q=2$ and $N_q=5$): 
there is no convergence in 
the statistics of the PCE approximation and for $\ph=3$ and $6$, respectively, the noise takes over and the 
results are meaningless. Table~\ref{tab:Ishigami2} shows that, using sufficient quadrature points, for an
increasing number of expansion terms the PCE variance converges to the data variance. If both variances are alike
also the Sobol' indices have converged to the true values. Still the number of expansion terms should not be taken too
large as can be seen for $\ph>9$ and $\ph>13$ where again the noise gradually takes over.

Finally, we also used the Saltelli method\cite{Saltelli02} - an improvement of the original Sobol' method - to compute the Sobol' indices for this problem. To reach a similar accuracy approximately 70-80 times as many sampling points are required, thus showing the gain in efficiency using the PCE-Gauss method to compute the Sobol' indices.

\subsection*{Software and computational dataset}
All software used in this paper is publicly available. The contact inhibition model resides at \url{http://sourceforge.net/projects/tst/}. For GSA we provide a repository containing the computational dataset and the 
analysis software at \url{http://persistent-identifier.org/?identifier=urn:nbn:nl:ui:18-23590}.

\section*{Results}
As a case study for the global sensitivity analysis (GSA) approach, we used a well-studied computational model of vascular morphogenesis: the contact inhibition model \cite{Merks2008}. We studied what single parameters and parameter interactions are important in the development of a spheroid of cells into vascular networks. For this purpose, we used the procedure outlined in Figure \ref{fig:workflow}: 1) select output measures, 2) select input parameters, 3) select a relevant subset of the global parameter space, 4) analyze the raw output, 5) perform GSA.

\subsection*{Selection of output measures}			
The contact inhibition model \cite{Merks2008} produces images of cell configurations as raw output. We chose two measures to quantify the raw output: compactness and lacuna count. Compactness of the network is a suitable measure of network development \cite{Merks2008} and is defined as the ratio $A_{\mathrm{cells}}/A_{\mathrm{hull}}$, with $A_{\mathrm{cells}}$ the number of lattice sites occupied by cells within a convex hull around all cells and $A_{\mathrm{hull}}$ the total number of lattice sites within the convex hull. A solid spheroid and a confluent monolayer of cells have a compactness close to one, while networks that contain lacunae have low values for compactness. Lacuna count is the number of lacunae in a network. Lacunae are defined as patches of medium (connected components of $\sigma=0$) enclosed by cells and are only counted when they have at least the size of a cell (50 lattice sites $\approx 200 ~\mu m^2$). 

\subsection*{Selection of input parameters}
The contact inhibition model \cite{Merks2008} is a stochastic, multi-factorial model. We refer to Section \textit{\nameref{methods}} for a detailed description of the model. The contact inhibition model has nine parameters: the number of cells ($N$), the target size of a cell ($A$), the rigidity of the cell ($\lambda_A$), cell-cell adhesion ($J_{\mathrm{cell},\mathrm{cell}}$), adhesion between cells and the extracellular matrix ($J_{\mathrm{cell},\mathrm{ECM}}$), the secretion rate of a chemoattractant by cells ($\alpha$), a diffusion coefficient of the chemoattractant ($D$), the decay rate of the chemoattractant ($\epsilon$), and a sensitivity of cells to the chemoattractant at cell-matrix interfaces ($\lambda_c$). 

In total, there are four model components or mechanisms in the contact inhibition model, namely cell size, adhesion, contact-inhibited chemotaxis and the gradient of the chemoattractant. In order to study the impact of each mechanism in the model extensively, we selected one parameter for each, ensuring that it is computationally feasible to generate enough data points for reliable GSA results. We thus selected four parameters: the cell rigidity ($\lambda_A$), cell-cell adhesion ($J_{\mathrm{cell},\mathrm{cell}}$), the diffusion coefficient of the chemoattractant ($D$), and a sensitivity of cells to the chemoattractant at cell-matrix interfaces ($\lambda_c$). The other parameters that regulate cell size ($A$), adhesion ($J_{\mathrm{cell},\mathrm{ECM}}$), or the gradient of the chemoattractant ($\epsilon$ and $\alpha$) will be fixed at the reference values corresponding to the values in \cite{Merks2008}. We kept the number of cells ($N$) in the spheroid constant, because we know from 
experience that it does not influence sprouting of spheroids in our model.

A GSA with four parameters can give new insights as four parameters are too many to obtain the relative impact of
the parameters and their interactions with visual plots or to know their effect solely by logic or intuition, while the number of simulations required for a GSA with four input parameters is computationally very feasible. A GSA with all parameters of the model is not expected to give additional information on the relative balance of the mechanisms and would be very time-consuming for a computationally intense model like the contact inhibition model. It would require roughly $10^9$ simulations (c.f., Section \textit{\nameref{method:reliable}}) to obtain reliable GSA results with all model parameters.

\begin{table}
\caption{Overview of the parameter selection for the GSA.
The names of the parameters are listed in the first column, a parameter description in the second column, and the selected parameter value ranges in the last column.}
      \begin{tabular}{ccc}
        \hline
         Name   & Description  & Range \\ \hline
        $\lambda_{c}$ & Chemotaxis  & 10 to 3000\\
         $D$  & Diffusion coefficient  & 1e-14 to 5e-13\\
        $\lambda_{A}$ & $\lambda$ Area  & 5 to 300\\
	$J_{\mathrm{cell},\mathrm{cell}}$ & Cell-cell adhesion  & 0 to 120\\\hline
      \end{tabular}
      \label{tab:par}
\end{table}

\subsection*{Selection of a relevant subset of the global parameter space}
To select the parameter ranges for which spheroids of cells develop into networks, we studied  one-dimensional parameter sweeps of the four selected input parameters for the compactness and lacuna count (Figure \ref{fig:1DSA}). The red lines in Figure \ref{fig:1DSA} represent the compactness and the blue lines the lacuna count. We selected the region in which the morphology of the network, and thus the value of the output measures, is changing and where no model artefacts arise. It is well studied for which parameter ranges artefacts arise in the CPM \cite{Anderson2007}, such as lattice anisotropy and `frozen' motility of cells. The regions shaded in gray indicate the deleted regions from the parameter space. For $\lambda_A$ the region 0 to 5 is deleted: cells cannot retain their volume here and disappear. This is a model artefact and does not represent a biological plausible situation. The region $\lambda_A>300$ is deleted, because 
cells are so rigid here that they hardly move. In the region $\lambda_c<10$ only spheroids form and for $\lambda_c>3000$ similar networks are always formed, thus these regions are deleted 
because the network morphology 
does not change. The parameters and their selected value ranges are listed in Table \ref{tab:par}. 

Based on the reliability study for the Ishigami test model (see Section \textit{\nameref{method:reliable}}), we expected that we required 10000 data points to perform a reliable GSA on our model. The points were chosen according to the Gauss-Legendre quadrature rule (see Section \textit{\nameref{method:GSA}}), resulting in ten values for each parameter. To correct for the stochasticity of the contact inhibition model, each parameter set is replicated twenty times with a different random seed and the output is averaged over them. The size of the standard deviations in Figure~\ref{fig:1DSA} indicate that the variation over different random seeds is very small for compactness, whereas the stochasticity in the model has a larger affect on the lacuna count. Nevertheless, this lacuna count is a reasonable measure for the network morphology.

 \begin{figure}
 \includegraphics[width=\textwidth]{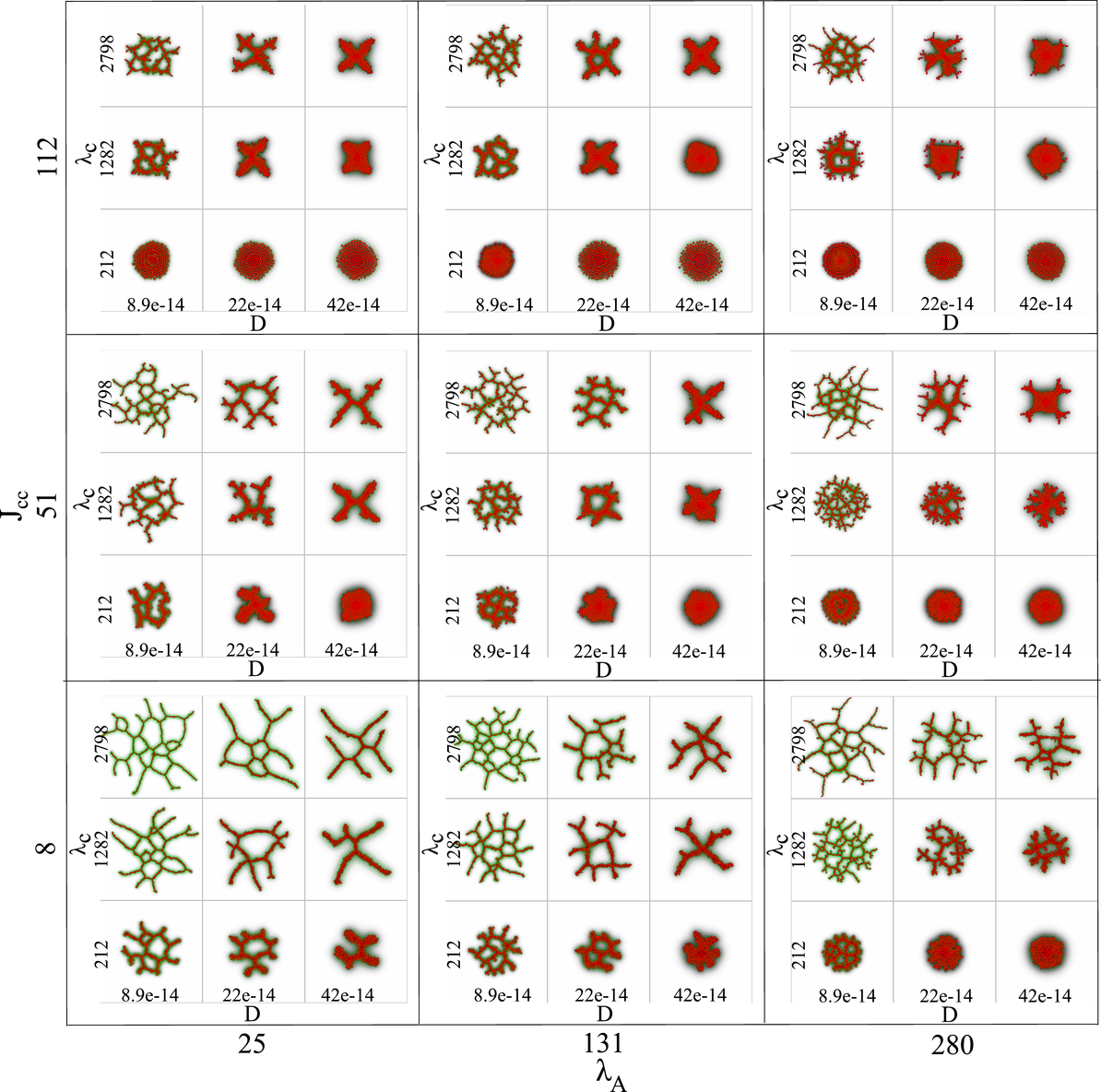}
  \caption{\textbf{Overview of the raw output.}
	An overview of the raw output of the contact inhibition model, the cell configurations at the end of a simulation, is shown in a collage of images. The cell rigidity ($\lambda_A$) is varied over the horizontal axis and cell-cell adhesion ($J_{\mathrm{cell},\mathrm{cell}}$) over the vertical axis. For each selected combination hereof, a subcollage is shown in which the diffusion coefficient of the chemoattractant ($D$) is plotted against the sensitivity of cells to the chemoattractant at cell-matrix interfaces ($\lambda_c$).
      }\label{fig:rawout}
	\end{figure}
	
\subsection*{Analysis of the raw output}
The raw output of a model simulation is an image of the cell configuration at the end of a simulation. Figure \ref{fig:rawout} gives an overview of the raw output for the selected parameter space. Examples of possible morphologies are shown in Figure \ref{fig:rawout}, ranging from spheroids to small networks with one lacuna, to fine-mazed networks with many lacunae. This is a visual reassurance that the input parameter space is well chosen. However, it is very difficult to predict from the raw output which parameters have a strong impact on the development of networks from spheroids. A GSA can give us insights into this, as we will show in the next section.

\subsection*{GSA of network development from spheroids}	
We performed two types of GSA on the distribution of the output measures, compactness and lacuna count, to study the impact of the parameters on vascular network development. The first type of GSA studies the variation of the output measures and the second type studies the decomposition of the variance of the distribution of the output measures.

\subsubsection*{GSA of the variation of the output measures}	
The variation in the output measures can be visualized by plotting the intensity of the output measures over two-dimensional slices of the parameter space. Figure \ref{fig:2DSALC} and \ref{fig:2DSACmp} show the intensity plots of the lacuna count and compactness, respectively, for each possible pairing of parameters. The parameter values are selected according to the Gauss-Legendre quadrature rule. 

Figure \ref{fig:2DSALC} shows that the diffusion coefficient is the main source of variation for the lacuna count: the lacuna count is high for low values of $D$ and low for high values of $D$, independent of the other parameters. That the lacuna count does not vary significantly over the entire perpendicular axis indicates that the parameter of the perpendicular axis does not have much impact. The dominance of the diffusion coefficient masks the impact of the other parameters. To reveal the impact of the other parameters, Additional file 4: Figure S1 caps the intensity values at a lacuna count of five.

Figure \ref{fig:2DSACmp} shows a high variation of the compactness in each plot. As a consequence, it is difficult to determine which parameters have a dominant impact on compactness. Interactions between parameters are difficult to deduce from these two-dimensional intensity plots. A variance-based GSA is well suited to derive parameter interactions and the ranking of individual parameter effects, as will be outlined in the following subsection.

\subsubsection*{Variance-based GSA of the output measures}	
To study the impact of single parameters and of parameter combinations on the development of networks from spheroids, we performed a GSA of the output distribution of compactness and lacunae count using the Sobol' indices. We refer to Section \textit{\nameref{method:GSA}} for a detailed description of how to obtain the 
Sobol' indices that represent the impact of the parameters. The GSA results of both measures are reliable, since the
Sobol' indices have converged for values of $\hat{p}$ for which $err_{\Var}$ (Equation \ref{eq:Errvar}) is small 
(see Additional file 1: Table S1 and Additional file 2: Table S2).

\begin{table}[h!]
\caption{Global sensitivity analysis results.
The Sobol' indices for the individual parameters (indices above mid-line) and for their combinations (indices below mid-line) are listed for the GSA of compactness and lacuna count.} 

      \begin{tabular}{lcc}
        \hline
										& compactness	&lacuna count\\\hline
				S($\lambda_{c}$)				&0.3188	&	0.0074\\
				S($D$)						&0.2969	&	0.7130	\\
				S($\lambda_{A}$)				&0.0266	&	0.0125	\\
				S($J_{\mathrm{cell},\mathrm{cell}}$)				&0.2048	&	0.0407	\\\hline
				S($\lambda_{c}$,$D$)				&0.0125	&	0.0570			\\
				S($\lambda_{c}$,$\lambda_{A}$)			&0.0107	&	0.0043	\\
				S($\lambda_{c}$,$J_{\mathrm{cell},\mathrm{cell}}$)		&0.0559	&	0.0145	\\
				S($D$,$\lambda_{A}$)				&0.0017	&	0.0347	\\
				S($D$,$J_{\mathrm{cell},\mathrm{cell}}$)				&0.0127	&	0.0521	\\
				S( $\lambda_{A}$,$J_{\mathrm{cell},\mathrm{cell}}$)		&0.0102	&	0.0048	\\
				S($\lambda_{c}$,$D$,$\lambda_{A}$)		&0.0102	&	0.0315	\\
				S($\lambda_{c}$,$D$,$J_{\mathrm{cell},\mathrm{cell}}$)		&0.0257	&	0.0232\\
				S($\lambda_{c}$,$\lambda_{A}$,$J_{\mathrm{cell},\mathrm{cell}}$)	&0.0217	&	0.0131\\
				S($D$,$\lambda_{A}$,$J_{\mathrm{cell},\mathrm{cell}}$)		&0.0075	&	0.0094\\\hline
			\end{tabular}
			\label{tab:sobol}
\end{table}

The second column of Table \ref{tab:sobol} lists the impact of the individual parameters and their combinations on compactness. The sensitivity for the chemoattractant at cell-matrix interfaces ($\lambda_c$) has the highest impact on network development (S($\lambda_c$)=0.3188), followed by the diffusion coefficient  with S($D$)=0.2969, and cell-cell adhesion with S($J_{\mathrm{cell},\mathrm{cell}}$)=0.2048. Elasticity of cells has a low impact of (S($\lambda_{A}$)=0.0266). Seventeen percent of the variance is caused by interactions of parameters. $\lambda_{c}$ and $J_{\mathrm{cell},\mathrm{cell}}$ have a combined impact of 0.0559. The impact of all other interactions was lower than S($\lambda_{A}$), which we will consider as a threshold for relevant impact.

The third column of Table \ref{tab:sobol} lists the impact of the individual parameters and their combinations on lacuna count. The individual impact of the diffusion coefficient is dominant, with S($D$)=0.7130. Cell adhesion also has a small individual impact (S($J_{\mathrm{cell},\mathrm{cell}}$)=0.0407). In total, twenty four percent of the variance is induced by combinations of parameters. There are five parameter combinations, which all include the diffusion coefficient, with a higher impact than the threshold: S($\lambda_{c}$,$D$)=0.0570, S($D$,$\lambda_{A}$)=0.0347, S($D$,$J_{\mathrm{cell},\mathrm{cell}}$)=0.0521, S($\lambda_{c}$,$D$,$J_{\mathrm{cell},\mathrm{cell}}$)=0.0476, and S($\lambda_{c}$,$D$,$\lambda_{A}$)=0.0315. The total impact of the diffusion coefficient is 90 percent. When we focus on low values of the lacuna count, by capping the lacuna count at a maximum of five lacunae, the dominance of the diffusion coefficient is slightly reduced and an extra interaction of 
$\lambda_{c}$ and $J_{\mathrm{cell},\mathrm{cell}}$ is 
found ($\lambda_{c}$,$J_{\mathrm{cell},\mathrm{cell}}$)=0.0409) (Additional file 3: Table S3).

	\begin{figure}
\includegraphics[width=\textwidth]{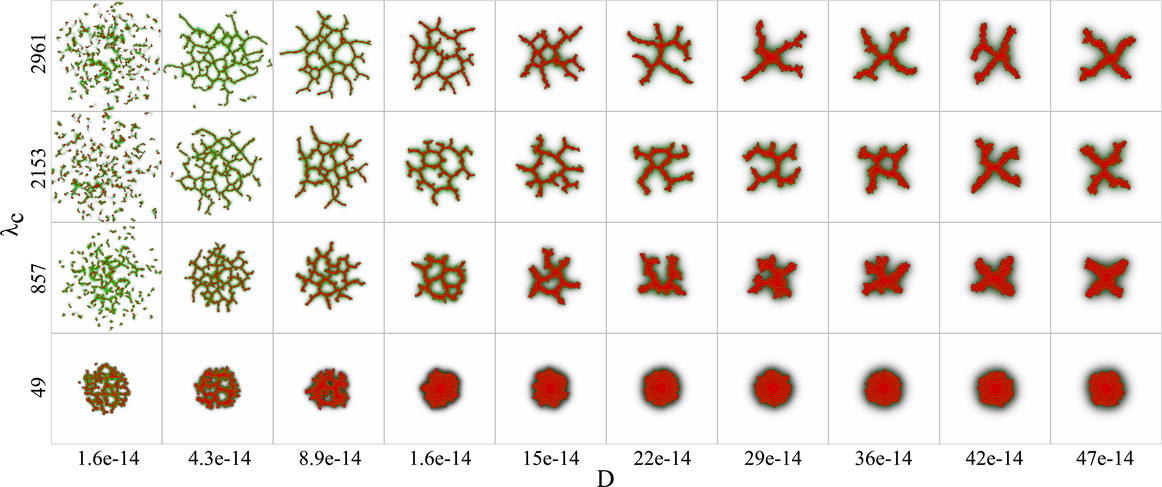}
  \caption{\textbf{Dominant effect of the diffusion coefficient on lacuna count.}
	A collage of the cell configurations at the end of simulations in the contact inhibition model, in which the diffusion coefficient of the chemoattractant ($D$) is varied over the horizontal axis and the sensitivity of cells to the chemoattractant at cell-matrix interfaces ($\lambda_c$) over the vertical axis.
      }
      \label{fig:diffusion}
      \end{figure}

\subsubsection*{Interpretation of the GSA results}	
The GSA results show that three parameters account for over 80 percent of the variance of the compactness distribution. Consistent with previous studies of the contact inhibition model \cite{Merks2008}, these three parameters are the diffusion coefficient, sensitivity to the chemoattractant at cell-matrix interfaces and cell-cell adhesion. For the lacuna count, the GSA identified solely the diffusion coefficient as the dominant parameter. This dominant effect is apparent in a collage of output images (Figure \ref{fig:diffusion}): the number of lacunae varies over the horizontal axis that represents the diffusion coefficient $D$, whereas there is little variation along the vertical axis that represents the sensitivity to the chemoattractant at cell-matrix interfaces $\lambda_c$. The number of lacunae is the largest for small values of the diffusion coefficient (around $D=4.3*10^{-14}$ $m^2/s$), whereas no lacunae are formed for large values of the diffusion coefficient. A similar trend is seen when the 
diffusion coefficient is plotted against cell-cell adhesion or cell rigidity (not shown). 
The distance over which adjacent branches can attract one another is given by the length of the chemoattractant gradient (Equation~\ref{eq:PDE}), which is characterized by the diffusion length, $L_{D}=\sqrt{\epsilon/D}$, the distance over which the secreted chemoattractant drops to $1/e$ of the concentration at the cells (see, {\it e.g.}, the discussion in Ref.~\cite{Merks2008}). If $L_{D}$ becomes shorter, branches that would fuse for larger values of $L_{D}$ will not fuse. Hence the pattern will be more fine-grained. Also a shorter value of $L_{D}$ will create sharper gradients and hence increase the inward chemotactic force (as $\Delta H = \lambda_c * \mathrm{gradient}$) hence "squeezing" the branches more and making them thinner. 

In conclusion, the GSA is able to identify the dominant single parameters for compactness and lacuna count. In addition, it gives new information on the relative ranking of the impact of these single parameters.		

In contrast to the one-dimensional parameter studies performed in \cite{Merks2008}, GSA provides information on interactions of parameters. Combinations of parameters account for 17\% of the variance in the compactness distribution, and for 24\% of the lacuna count distribution. This indicates that most parameters impact the model output independently. Interestingly, the parameter combination of $\lambda_{c}$ and $J_{\mathrm{cell},\mathrm{cell}}$ impacted the lacuna count (as seen in Additional file 4: Figure S1, which is capped at a maximum of five lacunae) as well as compactness. How can we explain this interaction? Sprout formation requires a balance between $\lambda_c$-dependent chemotaxis, creating an inward force perpendicular to the sprout surface, and $J_{\mathrm{cell},\mathrm{cell}}$-dependent cell-cell adhesion, which is responsible for the surface tension of individual cells. In the limit of zero-surface tension, the cells would behave as a zero-viscosity fluid and the chemotaxis would compress 
sprouts until they become infinitely thin \cite{ShirinifardThesis}. The cellular surface tensions resist such compression, thus determining the thickness of sprouts. Altogether, this parameter interdependence highlights a new insight in the mechanisms driving sprouting in our model.

\section*{Discussion}
Biological morphogenesis is a highly complicated process, involving genetic regulation, pattern formation, the biophysics of collective cell migration, mechanical cell-cell interactions, and so forth. As such multiscale mechanisms are practically impossible to understand intuitively, in recent years modeling and simulation has become a key tool in developmental biology (see, e.g., refs.~\cite{Sheth:2012db, DeRybel:2014ce, Besnard:2015kd, Buske:2012im}). These efforts have led to highly complicated models, where traditional analysis tools in dynamical systems theory, such as bifurcation analysis and phase plane analysis, fall short. The models must then be treated as `black-box' systems: one- or two-dimensional parameter sweeps are performed,  creating images and movies as output, which can be used to obtain various quantitative output measures. These parameter sweeps must be started from one or a few nominal parameter sets around which n-dimensional cross-shaped sweeps through the parameter space are 
performed. However insightful such studies are, a danger is that the impact of some parameters is overlooked: the conclusions may depend on what sets of nominal parameter values were selected. Using a simple, published simulation model of vascular morphogenesis, we have shown in this work how a multivariate GSA helps to get more insight in the relative impact of single parameters and of their interactions. We introduced a workflow for GSA of `black-box' models of morphogenesis. 

We applied the workflow to a vascular morphogenesis model which we refer to as the `contact inhibition model'. The output of the contact inhibition model consists of images of the cell configuration in a simulation. To quantify network development, we measured the compactness and the lacuna count of the cell configuration at the end of the simulation. A GSA with four input parameter distributions, that each described one of the four general model components, was performed for both measures. The GSA results of compactness and lacuna count both indicated that variation of the rigidity of the cells ($\lambda_A$) has very little impact on the model output. As a result, the model can be reduced by fixing this parameter. For compactness, the sensitivity for the chemoattractant at cell-matrix interfaces ($\lambda_c$) has the highest impact on network development, followed by the diffusion coefficient ($D$) and cell-cell adhesion ($J_{\mathrm{cell},\mathrm{cell}}$). In contrast, the GSA showed that the 
diffusion coefficient alone is dominant for lacuna count. The results for both measures are in line with what has been previously reported \cite{Merks2008}. New information from the GSA results is the relative impact of the single parameters. In addition, GSA identified interactions between parameters, which have led to new insights in the mechanism of sprouting in the model. Most notably, the parameter interactions in this specific CPM-based model have very low impact. Since GSA has not been performed for CPM-based models before, it is an important new insight for the CPM community that the most basic mechanisms of the CPM, such as cell size and adhesion here function independently.

Besides the contact inhibition model, there are multiple other computational models of vascular network development \cite{Oers2014,Szabo2010,Merks2006,KohnLuque2011,Bauer2007,Shirinifard2009}. These models often share common mechanisms that drive sprouting, but differ by one or a few mechanisms. It is still not known which mechanisms drive sprouting \emph{in vivo}, or whether a different set of mechanisms is used under different conditions. We propose GSA as an approach to help falsify these models. Firstly, the ranking of the relevance of the mechanisms in the models can be compared with knowledge of the impact of these mechanisms from experimental data to falsify models. A second model falsification method is the validation of the experimental 
predictions of each model based on the GSA results. 

The workflow is designed to take into account some pitfalls of GSA. These arise from the dependency of the outcome on  the choices one makes for the output measure, input parameters and their distributions. Different output measures can give different results, as was the case for compactness and lacuna count. This indicates that it is essential to consider carefully whether the selected measure truly describes your goal and if there are other measures for it. A selection of input parameters might be necessary when it is not computationally feasible or methodologically desirable to use all parameters of the model. The importance of the selection of the correct parameter distributions has also been discussed elsewhere \cite{GSABangaDoyle12}. Intuitively, a large range for the parameter values will allow for the largest variation in the output and thus the most interesting result. However, since the analysis is global over the entire parameter space, local though important features might become unnoticed if the 
distribution is too widespread. For instance, for the contact inhibition model we were interested in the region where the networks developed and where the measures were changing accordingly, and variation in these regions could become unnoticed if we included large regions where for instance spheroids do not sprout. Ideally, the parameter distribution comes from experimental measurements, but in absence hereof we propose to study the variation of the output measures for each parameter 
individually.

It is crucial to have an estimate of the accuracy of the sensitivity results.
One option is to compare the results with the outcome of an analysis with a
higher accuracy computed with more quadrature points and a higher PCE order,
like advocated in \cite{GSABangaDoyle12}. In this paper we proposed a simpler
and cheaper rule: given the number of quadrature points the Sobol' indices
should show convergence for those values of $\ph$ for which the variance computed with
the Gauss-Legendre quadrature rule is more or less equal to the variance
computed from the PCE approximation. If a higher PCE order is required, more
model simulations are needed. Since the computation of the PCE statistics is
`for free' compared to model simulations this is an efficient way of judging
whether the accuracy of the statistics is sufficient for one's aim.
Although Gauss quadrature is optimal, it has the disadvantage that it is not
a nested quadrature rule, i.e., if more quadrature points are required, the
old model results cannot be re-used.
An alternative for Gauss quadrature is Monte Carlo (MC) integration.
Sampling the PCE integrals by MC is less optimal, so more simulations are
needed to obtain reliable GSA results. For the Ishigami test model, MC needs a 100 times more simulations to get comparable results. The benefit of MC is that you can check
`on the fly' if there are enough data points generated to get reliable results.
Adding simulations on the fly is particular useful when the estimated number
of simulations based on the Gauss quadrature rule is computationally unfeasible,
but one expects or hopes that the output distribution is relatively smooth and
thus can be described by a low order PCE approximation.

Some studies require GSA of a subspace of the output distribution. For instance in our case study, to study not the conditions for network formation per se, but the details of the network morphology (e.g. branch length, branch thickness, and so forth), we must preselect a region of the parameter space where networks actually form. Unfortunately, such a subspace would no longer guarantee that the input distribution is independently random uniform. For such cases, a more complicated method to compute the Sobol' indices \cite{gPCcor14} is required. 

Besides in computational models, the impact of biological factors on morphogenesis is also studied \emph{in vitro}. High-throughput image-based screenings systematically analyze the impact of genes or potential drugs on cell behavior, such as cell migration \cite{Devedec2010}. This `systems microscopy' approach is well suited for parallel screening of cellular responses to numerous experimental perturbations \cite{Lock2010}.  Such high-throughput screens can be performed for the genes, growth factors or ECM concentrations affecting morphogenesis. This is conceptually very similar to parameter studies of \emph{in silico} `black-box' models. The perturbed biological factors represent the input parameters and the output is an image from which quantitative data can be derived. Therefore, the GSA workflow proposed in this paper is directly applicable to high-throughput \emph{in vitro} studies.

\section*{Conclusions}
Morphogenesis is a complex biological process in which cells organize into shapes and patterns. Computational modeling is used to get insights in the mechanisms of morphogenesis. These models are often multi-scale, non-linear and multi-factorial, making it difficult to relate their input to their output. The behavior of such `black-box' models is mostly studied by visual inspection and analyses of the individual output (e.g. images and movies) and with one- or two-dimensional parameter sweeps of output measures. However, this does not provide insight in the relative impact of single parameters and of their interactions on the outcome of the model. GSA fulfills this task. GSA results can give insights in the dynamics of the model and help to generate experimental predictions to manipulate morphogenesis. In this paper, we introduced a workflow for GSA of such models and addressed pitfalls and reliability of the analysis. The workflow is applied to the contact inhibition model, a cell-based model of vascular 
morphogenesis. 
GSA was 
able to correctly identify dominant parameters and gave new insights on the magnitude and ranking of their individual impact and importantly, on their interactions. In summary, we propose GSA of `black-box' models, such as complex computational models or high-throughput \textit{in vitro} models, as an alternative approach to get insights in the mechanisms of morphogenesis.

\section*{Abbreviations}
\begin{itemize}
 \item[CPM] cellular Potts model
 \item[ECM] extracellular matrix
 \item[GSA] global sensitivity analysis
 \item[MC] Monte Carlo
 \item[MCS] Monte Carlo step
 \item[PCE] polynomial chaos expansion
 \item[PDE] partial differential equation
 \item[pdf] probability density function
 \item[VEGF]vascular endothelial growth factor
\end{itemize}
%FAST
%MLR
%DGSM

\section*{Competing interests}
The authors declare that they have no competing interests.

\section*{Author's contributions}
SB, MNJ, RM and JB conceived of the study. SB performed the computational simulations, participated in the data analysis and drafted the manuscript. MNJ analyzed the data and helped to draft the manuscript. RM participated in the design and coordination of the study and helped to draft the manuscript. JB participated in the design and coordination of the study and in the data analysis, and helped to draft the manuscript. All authors read and approved the final manuscript.

\section*{Acknowledgements}
We thank Indiana University and the Biocomplexity Institute for providing the CompuCell3D
modeling environment and SURFsara (www.surfsara.nl) for the support in using the Lisa
Compute Cluster.

%\section*{Funding statement}
The investigations were supported by the Division for Earth and Life Sciences (ALW) with
financial aid from the Netherlands Organization for Scientific Research (NWO).

%%  After submission of the .TEX file,                     %%
%%  you will be prompted to submit your .BBL file.         %%
%%                                                         %%
%%                                                        \hline         %%

% if your bibliography is in bibtex format, use those commands:
%\bibliographystyle{unsrt} % Style BST file
%\bibliography{bmc_article}      % Bibliography file (usually '*.bib' )

\newpage
\section*{Supplementary Information}

\subsection*{Additional file 1 --- Global sensitivity analysis results for compactness}

\textbf{Table S1.} Global sensitivity analysis results for compactness.

      \begin{tabular}{lcccc}
        \hline
				$\hat{p}$	&	12	&	13	&	14	&	15	\\
				Variance  data	&	0.0453	&	0.0453	&	0.0453	&	0.0453	\\
				Variance PCE																	&	0.0452	&	0.0453	&	0.0454	&	0.0456	\\
				S($\lambda_{c}$)														&	0.3190	&	0.3188	&	0.3186	&		0.3183\\
				S($D$)																			&	0.2971	&	0.2969	&	0.2965	&	0.2958	\\
				S($\lambda_{A}$)														&	0.0267&		0.0266	&	0.0266	&	0.0265	\\
				S($J_{\mathrm{cell},\mathrm{cell}}$)													&	0.2052	&	0.2048	&	0.2043	&	0.2032	\\
				S($\lambda_{c}$,$D$)								&	0.0124	&	0.0125	&	0.0126	&0.0127		\\
				S($\lambda_{c}$,$\lambda_{A}$)						&	0.0107	&	0.0107	&	0.0109	&	0.0110	\\
				S($\lambda_{c}$,$J_{\mathrm{cell},\mathrm{cell}}$)					&	0.0558	&	0.0559	&	0.0561&	0.0570	\\
				S($D$,$\lambda_{A}$)								&	0.0016	&	0.0017	&	0.0017	&0.0017		\\
				S($D$,$J_{\mathrm{cell},\mathrm{cell}}$)							&	0.0127	&	0.0127	&	0.0127	&	0.0126	\\
				S( $\lambda_{A}$,$J_{\mathrm{cell},\mathrm{cell}}$)					&	0.0102	&	0.0102	&	0.0102	&0.0101	\\
				S($\lambda_{c}$,$D$,$\lambda_{A}$)	&	0.0098	&	0.0102	&	0.0104	&	0.0107	\\
				S($\lambda_{c}$,$D$,$J_{\mathrm{cell},\mathrm{cell}}$)	&	0.0253	&	0.0257	&	0.0262	&	0.0268	\\
				S($\lambda_{c}$,$\lambda_{A}$,$J_{\mathrm{cell},\mathrm{cell}}$)	&	0.0214	&0.0217		&	0.0220	&	0.0225	\\
				S($D$,$\lambda_{A}$,$J_{\mathrm{cell},\mathrm{cell}}$)	&		0.0073&	0.0075	&	0.0076	&	0.0078	\\\hline
			\end{tabular}
			\label{table:Sobolcomp}

\subsection*{Additional file 2 --- Global sensitivity analysis results for lacuna count}
\textbf{Table S2.} Global sensitivity analysis results for lacuna count.

      \begin{tabular}{lccc}
        \hline
				$\hat{p}$				&	11	&	12	&	13	\\
				Variance  data				&	15.8651	&	15.8651	&	15.8651	\\
				Variance PCE				&	15.5295	&	15.9661	&	16.7974	\\
				S($\lambda_{c}$)			&	0.0075	&	0.0074	&	0.0070	\\
				S($D$)					&	0.7120	&	0.7130	&	0.7187	\\
				S($\lambda_{A}$)			&	0.0129	&	0.0125	&	0.0119	\\
				S($J_{\mathrm{cell},\mathrm{cell}}$)			&	0.0418	&	0.0407	&	0.0387	\\
				S($\lambda_{c}$,$D$)			&	0.0579	&	0.0570	&	0.0552	\\
				S($\lambda_{c}$,$\lambda_{A}$)		&	0.0043	&	0.0043	&	0.0041	\\
				S($\lambda_{c}$,$J_{\mathrm{cell},\mathrm{cell}}$)	&	0.0144	&	0.0145	& 	0.0143	\\
				S($D$,$\lambda_{A}$)			&	0.0350	&	0.0347	&	0.0348	\\
				S($D$,$J_{\mathrm{cell},\mathrm{cell}}$)			&	0.0533	&	0.0521	&	0.0501	\\
				S( $\lambda_{A}$,$J_{\mathrm{cell},\mathrm{cell}}$)	&	0.0050	&	0.0048	&	0.0046	\\
				S($\lambda_{c}$,$D$,$\lambda_{A}$)	&	0.0289	&	0.0315	&	0.0331	\\
				S($\lambda_{c}$,$D$,$J_{\mathrm{cell},\mathrm{cell}}$)	&	0.0222	&	0.0232	&	0.0236	\\
				S($\lambda_{c}$,$\lambda_{A}$,$J_{\mathrm{cell},\mathrm{cell}}$)& 0.0118	&	0.0131	&	0.0142	\\
				S($D$,$\lambda_{A}$,$J_{\mathrm{cell},\mathrm{cell}}$)	&	0.0086	&	0.0094	&	0.0100	\\\hline
			\end{tabular}

\subsection*{Additional file 3 --- Global sensitivity analysis results for lacuna count capped at maximum of 5 lacunae}
\textbf{Table S3} Global sensitivity analysis results for lacuna count capped at maximum of 5 lacunae. 

      \begin{tabular}{lccc}
        \hline
				$\hat{p}$					&	12	&	13	&	14	\\
				Variance  data					&	3.4870	&	3.4870	&	3.4870	\\
				Variance PCE					&	3.4346	&	3.4821	&	3.5668	\\
				S($\lambda_{c}$)				&	0.0081	&	0.0081	&	0.0079	\\
				S($D$)						&	0.6893	&	0.6861	&	0.6847	\\
				S($\lambda_{A}$)				&	0.0078	&	0.0077	&	0.0077	\\
				S($J_{\mathrm{cell},\mathrm{cell}}$)				&	0.0578	&	0.0571	&	0.0557	\\
				S($\lambda_{c}$,$D$)				&	0.0561	&	0.0557	&	0.0547	\\
				S($\lambda_{c}$,$\lambda_{A}$)			&	0.0084	&	0.0085	&	0.0086	\\
				S($\lambda_{c}$,$J_{\mathrm{cell},\mathrm{cell}}$)		&	0.0394	&	0.0409	&	0.0429	\\
				S($D$,$\lambda_{A}$)				&	0.0048	&	0.0048	&	0.0048	\\
				S($D$,$J_{\mathrm{cell},\mathrm{cell}}$)				&	0.0489	&	0.0485	&	0.0476	\\
				S( $\lambda_{A}$,$J_{\mathrm{cell},\mathrm{cell}}$)		&	0.0035	&	0.0035	&	0.0035	\\
				S($\lambda_{c}$,$D$,$\lambda_{A}$)		&	0.0147	&	0.0160	&	0.0169	\\
				S($\lambda_{c}$,$D$,$J_{\mathrm{cell},\mathrm{cell}}$)		&	0.0458	&	0.0476	&	0.0490	\\
				S($\lambda_{c}$,$\lambda_{A}$,$J_{\mathrm{cell},\mathrm{cell}}$)	&	0.0239	&	0.0258	&	0.0274	\\
				S($D$,$\lambda_{A}$,$J_{\mathrm{cell},\mathrm{cell}}$)		&	0.0121	&	0.0132	&	0.0140	\\\hline
			\end{tabular}

\subsection*{Additional file 4 --- Two-dimensional intensity plots of lacuna count with maximum intensity of 5 lacunae}

\includegraphics[width=\textwidth]{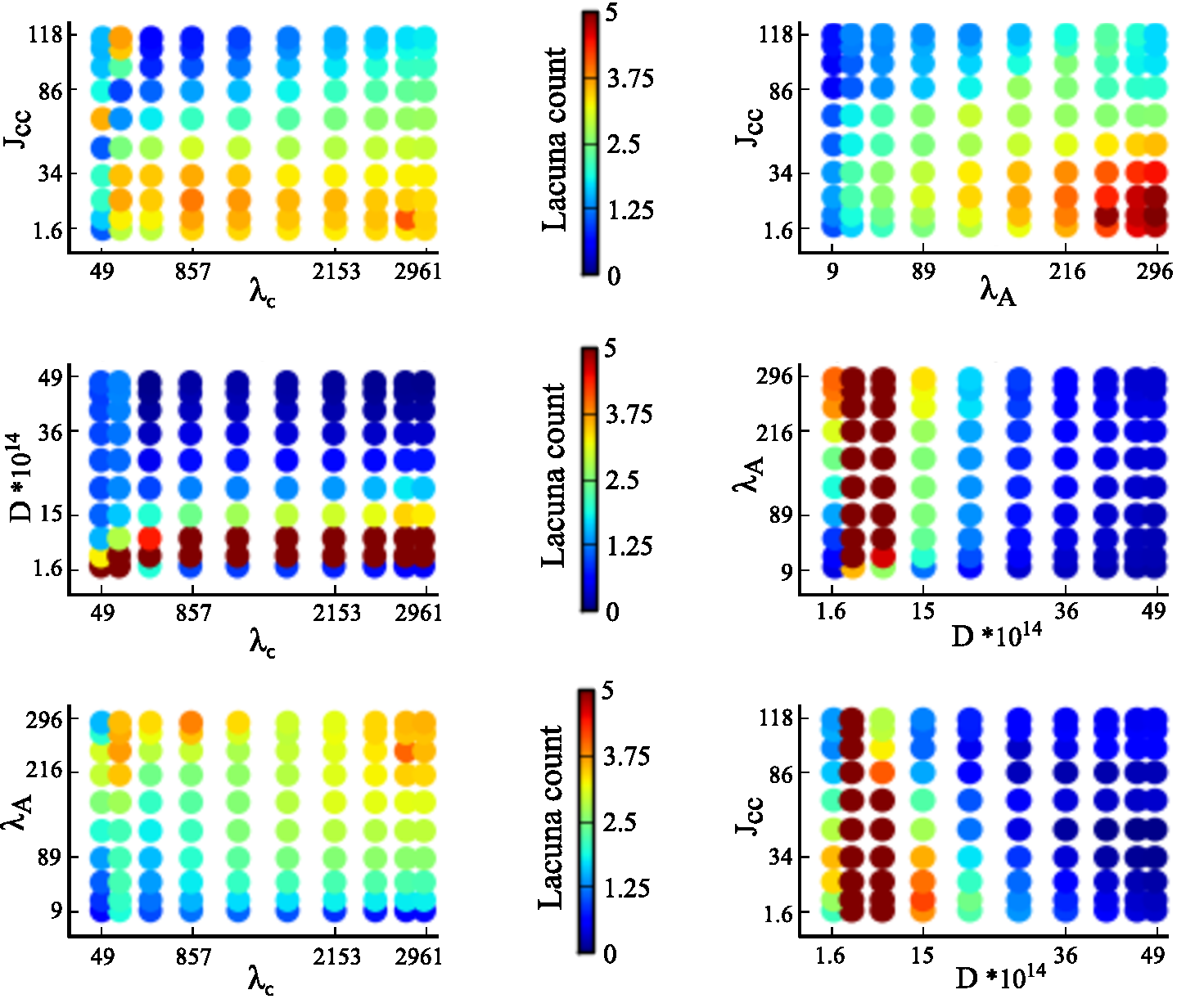}
 \textbf{Figure S1. Two-dimensional intensity plots of lacuna count with maximum intensity of 5} The intensity of the output measure lacuna count, mapped to an interval of 0 to 5 as indicated by the color bars, is plotted for each two-parameter combination of the parameters the cell rigidity ($\lambda_A$), cell-cell adhesion ($J_{\mathrm{cell},\mathrm{cell}}$), the diffusion coefficient of the chemoattractant ($D$), and sensitivity of cells to the chemoattractant at cell-matrix interfaces ($\lambda_c$).

\end{document}